\documentstyle[12pt]{article}\textheight
240mm\textwidth 170mm \hoffset=-1.5cm
\newcommand{\bi}{\bibitem}
\newcommand{\be}{\begin{eqnarray}}
\newcommand{\ee}{\end{eqnarray}}
\newcommand{\nn}{\nonumber}
\catcode`\@=11
\def\lsim{\mathrel{\mathpalette\@versim<}}
\def\gsim{\mathrel{\mathpalette\@versim>}}
\def\@versim#1#2{\vcenter{\offinterlineskip
\ialign{$\m@th#1\hfil##\hfil$\crcr#2\crcr\sim\crcr } }}
\catcode`\@=12

\begin{document}
\pagestyle{empty}
\noindent
\hspace*{10.7cm} \vspace{-3mm} DPSU-97-3\\
\hspace*{10.7cm} \vspace{-3mm} INS-Rep-1186\\
\hspace*{10.7cm} \vspace{-3mm} KANAZAWA-97-02\\

\hspace*{10.7cm} March 1997

\vspace{0.3cm}

\begin{center}
{\Large\bf  
 Soft Scalar-Mass Sum Rule
in Gauge-Yukawa Unified Models
and Its Superstring Interpretation}
\end{center} 

\vspace{1cm}

\begin{center}{\sc Yoshiharu Kawamura}$\ ^{(1)}$, 
{\sc Tatsuo Kobayashi}$\ ^{(2)}$ and 
{\sc Jisuke Kubo}$\ ^{(3),*}$
\end{center}
\begin{center}
{\em $\ ^{(1)}$  Department of Physics, Shinshu University, \\
Matsumoto, 390 Japan }\\
{\em $\ ^{(2)}$ Institute for Nuclear Study, University of Tokyo \\
   Midori-cho, Tanashi, Tokyo, 188 Japan} \\
{\em $\ ^{(3)}$ 
Physics Department, \vspace{-2mm} Faculty of Science, \\
Kanazawa \vspace{-2mm} University,
Kanazawa, 920-11 Japan } 
\end{center}

\vspace{1cm}
\begin{center}
{\sc\large Abstract}
\end{center}

\noindent
It is shown that
a certain sum rule for soft supersymmetry--breaking
scalar masses, \vspace{-2mm} which has been recently found
in a certain class of  superstring models, \vspace{-2mm} is universal for
gauge-Yukawa unified models. 
To explain this coincidence, we argue that 
the low-energy remnant \vspace{-2mm} of the target-space duality
invariance  in the effective \vspace{-2mm}
supergravity of compactified superstrings
can be identified  with the (broken) scale invariance
in gauge-Yukawa unified models, \vspace{-2mm} and
that gauge-Yukawa unification which is 
indispensable for the sum rule to be satisfied  \vspace{-2mm} 
follows from the matching of anomalies.

\vspace*{3cm}
\footnoterule
\vspace*{2mm}
\noindent
$^{*}$Partially supported  by the Grants-in-Aid
for \vspace{-3mm} Scientific Research  from the Ministry of
Education, Science 
and Culture \vspace{-3mm}  (No. 40211213).\\

\newpage
\pagestyle{plain}
Gauge-Yukawa Unification 
(GYU) \cite{kmz1,kubo1}  is an
attempt to relate  the gauge and Yukawa couplings, which is 
a gradual  extension of the Grand Unification idea. 
It has turned out to be a successful scheme to  predict
 the top and bottom quarks masses \cite{kmoz2}.
Supersymmetry seems  to be essential for GYU,
but, as it is for any realistic  supersymmetric model,
the breaking of supersymmetry
has to be understood.
If a model couples to
 supergravity, 
or in the case of 
 gauge--mediated supersymmetry breaking 
\footnote{See refs. \cite{nilles1}, \cite{dine1} 
and references therein.}, one can compute in
principle  the soft supersymmetry--breaking (SSB) terms. 
Unlike this usual path chosen to reduce the number of 
the independent parameters,
the GYU idea of \cite{kmz1,kubo1} relies  not only  
on a symmetry principle, but also on the principle of
reduction of couplings \cite{zim1,kubo2}. This  principle is based on the
existence of renormalization group
(RG) invariant relations among couplings, which 
do not necessarily result from a symmetry,
but nevertheless preserve perturbative renormalizability. 
Dimensional couplings can also be treated
along this line of 
thought \cite{piguet1}--\cite{kubo3}. When applied to
 the finite \cite{jack1,kazakov1}  or  the minimal  \cite{kubo3} 
supersymmetric $SU(5)$ GUT, one finds that
the SSB sector of the model
is  completely fixed by the gaugino mass $M$.

One of  main observations of this letter is that the
soft scalar masses in GYU models satisfy  
a  simple universal sum rule 
eq. (\ref{sumr}). This sum rule is derived without using any
symmetry of  GYU models. Therefore, within the framework
of  GYU,
the sum rule is an accidental byproduct,  but its simplicity
suggests that it could be understood as a consequence of some
symmetry property of a more fundamental theory such as 
superstrings.
Superstring theory, though intensive studies and recent
interesting developments \cite{string}, is still in a phase in which
one needs various assumptions,
especially those on
non-perturbative effects, to relate it 
to low energy physics and then to predict
its parameters. These assumptions include also  those on supersymmetry
breaking. 
Nevertheless,
it is possible to do  systematic investigations
of SSB terms \cite{SSB}
 and to parametrize them in 
a simple way \cite{BIM}--\cite{kim1} (see ref. \cite{ibanez1} for a review).
Along this line, we re-investigate the K\" ahler potential under general
assumptions to find out its general form that yields the  
 sum rule which coincides with the one in GYU models.
As expected,  the sum rule results from
a certain type of duality invariance  
 (see refs. 
\cite{ferrara1,IL,S-dual} and
references  therein) in
the effective supergravity.
In fact, the same sum rule has been independently
obtained in various superstring 
models \cite{BIM,BIM2,kim1}.

As we will also see, the unification of the gauge and
Yukawa couplings 
 is indispensable for the sum rule to be 
satisfied. This appears  mysterious,
because one can derive the sum rule in superstrings without
explicitly knowing the superpotential.
To give a possible answer to this problem, we notice
that the duality transformation acts as field-dependent
scale transformations on the chiral matter
superfields. Given that, we may identify
the low-energy remnant of the duality invariance
with the (in general broken) scale invariance of the effective
renormalizable field theory.
We will argue that
this interpretation might offer a possibility to understand
 why in  GYU models and in
the certain class of superstring 
models \cite{BIM,BIM2,kim1}
the same  soft scalar-mass sum rule is satisfied,
at least in one-loop order, to  which we will be restricting
ourselves throughout this letter.

Let us first derive the announced  soft scalar-mass sum rule.
To this end,   we 
 use the notation and result of ref. \cite{martin1}.
The superpotential (the gauge group is assumed to be a simple group) 
is given by 
\be
W &=&\frac{1}{6} \,Y^{ijk}\,\Phi_i \Phi_j \Phi_k
+\frac{1}{2} \,\mu^{ij}\,\Phi_i \Phi_j~,
\ee
along with the Lagrangian for SSB terms,
\be
-{\cal L}_{\rm SB} &=&
\frac{1}{6} \,h^{ijk}\,\phi_i \phi_j \phi_k
+
\frac{1}{2} \,b^{ij}\,\phi_i \phi_j
+
\frac{1}{2} \,(m^2)^{j}_{i}\,\phi^{*\,i} \phi_j+
\frac{1}{2} \,M\,\lambda \lambda+\mbox{H.c.}~
\ee
The  RG functions we need for our purpose are:
\be
\frac{d}{dt}\,g &=& \beta_{g} = 
\frac{1}{16\pi^2}\beta_{g}^{(1)}+\dots~,~
\frac{d}{dt}\,M = \beta_{M} = 
\frac{1}{16\pi^2}\beta_{M}^{(1)}+\dots~,\\
\frac{d}{dt}\,Y^{ijk} &=& \beta_{Y}^{ijk}=Y^{ijp}\,\{
~\frac{1}{16\pi^2}\,\gamma_{p}^{(1)\,k} +\dots~ \}+(k \leftrightarrow i)
+(k\leftrightarrow j)~,\\
\frac{d}{dt}\,h^{ijk} &= &\beta_{h}^{ijk}=\frac{1}{16\pi^2}\,
[\beta_{h}^{(1)}]^{ijk}+\dots ~,~
\frac{d}{dt}\,(m^2)^{j}_{i} =[\beta_{m^2}]^{j}_{i}=\frac{1}{16\pi^2}\,
[\beta_{m^2}^{(1)}]^{j}_{i}+\dots~,
\ee
where  $\dots$ stands for higher order terms (see ref.  \cite{martin1} and
references therein)
\be
\beta_{g}^{(1)} & = & g^3\,[S(R)-3 C(G)]~,~
\beta_{M}^{(1)}=2 M\, \beta_{g}^{(1)}/g~,~
\gamma_{i}^{(1)\,j} = (1/2) Y_{ipq} Y^{jpq}
-2\delta^{j}_{i}\, g^2\,C(i)~, \label{beta1}
\ee
\be
[\beta_{h}^{(1)}]^{ijk} &=& (1/2) h^{ijl} Y_{lmn} Y^{mnk}+
 Y^{ijl} Y_{lmn} h^{mnk}-2(h^{ijk}-2M Y^{ijk})\,g^2\,C(k) \nn\\
& &+(k \leftrightarrow i)+(k \leftrightarrow j)~,
\ee
\be
[\beta_{m^2}^{(1)}]^{j}_{i} &=& (1/2) Y_{ipq} Y^{pqn} (m^2)_{n}^{j}+
(1/2) Y^{jpq} Y_{pqn} (m^2)_{i}^{n}+
2Y_{ipq}Y^{jpr}(m^2)_{r}^{q} \nn\\
& &+h_{ipq} h^{jpq} -8\delta_{i}^{j} M M^{\dag}\,g^2\,C(i)~,
\ee
The soft scalar-mass sum rule which we would like to derive is given by
\be
m_{i}^{2}+m_{j}^{2}+m_{k}^{2}
 &=&M M^{\dag}~~\mbox{for}~~i,j,k~~\mbox{with}~\rho^{ijk} \neq 0~,
\label{sumr}
\ee
where $\rho^{ijk}$ is defined in eq. (\ref{Yg}) below.
(Each set of the
subscripts  $\{ i,j,k \}$  appearing in the sum rule exactly  refers to a
non-vanishing cubic term in the superpotential.)
The
sum rule (\ref{sumr}) is satisfied under two conditions
which we will  specify at the corresponding points below.
These conditions are by no means strong, and indeed they are satisfied in
all the known GYU models so far.
Note that the sum rule (\ref{sumr}) is a one-loop result, and so 
we expect it will be modified in higher orders in perturbation theory.

We proceed from the starting assumption that
 the Yukawa couplings $Y^{ijk}$  are 
expressed in terms  of the gauge coupling $g$:
\be 
Y^{ijk} &=& \rho^{ijk}g+\dots~,
\label{Yg}
\ee
where $\rho^{ijk}$ are constants independent of
$g$ and   $\dots$ stands for higher order terms.
Eq. (\ref{Yg}) is the one-loop solution to the reduction 
equation \cite{zim1} 
\be
\beta_{Y}^{ijk} &=& \beta_{g}\,d Y^{ijk}/d g~.
\label{Yg2}
\ee
In the case of a finite
theory, the $\beta$ functions vanish, of course.
Nevertheless, the reduction equation keeps its meaning
\cite{zim1,LPS}, and as we will see later on, the sum rule (\ref{sumr})
 can be derived  for that case, too. We however
assume for a while that the $\beta$ functions do not vanish.

\vspace{0.2cm}
\noindent
{\bf Condition I}

The coefficients $\rho^{ijk}$ 
satisfy the diagonality relation
\be
\rho_{ipq}\rho^{jpq} &\propto & \delta_{i}^{j}~.
\label{cond1}
\ee

\vspace{0.2cm}
\noindent
This condition implies that the one-loop 
anomalous dimensions for $\Phi_i$'s
become diagonal if the reduction solution (\ref{Yg})
 is inserted, i.e., $\gamma_{i}^{(1)\,j} = 
\gamma_{i}\,\delta^{j}_{i}\,g^2 $,
where $ \gamma_{i}$ are also constants independent of
$g$.
Therefore, the one-loop
$\beta$ function for $Y^{ijk}$ in the reduced theory takes the form
\be
[\beta_{Y}^{(1)}]^{ijk}/16\pi^2 &=& \rho^{ijk}\,(\gamma_{i}+
\gamma_{j}+\gamma_{k}) \,g^3 /16\pi^2~.
\label{betaY}
\ee
The reduction equation (\ref{Yg2}), furthermore, requires that
\be
\sum_l {}^{\rho}\gamma_l &\equiv &
\gamma_{i}+\gamma_{j}+\gamma_{k} = 
\beta_{g}^{(1)}/g^3=
S(R)-3C(G)
\label{sumg} 
\ee
for $\{ i,j,k \}$ with $\rho^{ijk} \neq 0$.
(This implies that all  the allowed cubic coupling terms 
$Y^{ijk}\Phi_i \Phi_j \Phi_k$ transform in the same way under the 
 scale transformation.)
Note further that
\be
h^{ijk} &=& -M Y^{ijk}+\dots =-M \rho^{ijk}\,g+\dots
\label{hY}
\ee
solves the reduction equation for $h^{ijk}$ \cite{kubo3},
\be
\beta_{h}^{ijk} &=&\beta_{M}\partial h^{ijk}/\partial M
+\beta_{M^{\dag}}\partial  h^{ijk}/\partial M^{\dag}
+\beta_{g}\partial h^{ijk}/\partial g~,
\label{redh}
\ee
in one-loop order.
This can be shown from
$ [\beta_{h}^{(1)}]^{ijk} =
-3 [\beta_{Y}^{(1)}]^{ijk} M$,
which is a consequence of 
(\ref{hY}) \footnote{Similarly, we  can obtain the 
reduction solution for the $B$-term
(under a certain assumption),
$ b^{ij}=-(\gamma_i+\gamma_j)/(S(R)-3C(G)) \mu^{ij}M $,
which leads to another type of sum rule,
$ m_i^2+m_j^2+ (b^{ij}/ \mu^{ij}) M^{\dag} = 0 $.}.
 
\vspace{0.2cm}
\noindent
{\bf Condition II}

The one-loop reduction solution for the scalar masses
is diagonal, i.e., 
\be
(m^2)^{j}_{i} &=& m^{2}_{i}\,\delta^{j}_{i}~,~
 m^{2}_{i} = \kappa_{i} M M^{\dag}~,
\label{cond2}
\ee
where $\kappa_i$ are constants to be determined below.

\vspace{0.2cm}
\noindent
If the soft scalar-mass matrix is  diagonal, the one-loop
$\beta$ functions $[\beta_{m^2}^{(1)}]^{j}_{i}$ can be written as
\be
[\beta_{m^2}^{(1)}]^{j}_{i} &=& \rho_{ipq}
\rho^{jpq}\,(~m_{i}^{2}/2+m_{j}^{2}/2+m_{p}^{2}+m_{q}^{2}~)g^2
+h_{ipq} h^{jpq}-8\delta^{j}_{i} \,M M^{\dag}g^2 C(i)  
\label{betam} \\
 & & \propto
\delta^{j}_{i}~.\nn
\ee
Using eqs. (\ref{cond1}) and (\ref{hY}) we then see that  
$\rho_{ipq} \rho^{jpq}\,(m_{p}^{2}+m_{q}^{2}~)$ also has to
be proportional to
$\delta^{j}_{i}$. This implies that,
if the sum rule (\ref{sumr}) is satisfied, 
the r.h.s of eq. (\ref{betam}) becomes
\be
 \{~2\rho_{ipq} \rho^{jpq}-8
\delta^{j}_{i} C(i)~\} M M^{\dag}\,g^2
=4\gamma_{i} \,\delta^{j}_{i}\,M M^{\dag}\,g^2~,
\ee
where we have used eqs. (\ref{beta1}) and (\ref{betaY}). From 
the reduction equation for $(m^2)^{j}_{i}$,
\be
[\beta_{m^2}]^{j}_{i} &=&\beta_{M}
\partial (m^2)^{j}_{i}/\partial M+
\beta_{M^{\dag}}\partial (m^2)^{j}_{i}/\partial M^{\dag}
+\beta_{g}\partial (m^2)^{j}_{i}/\partial g~,
\label{redm}
\ee
we finally obtain
\be
\kappa_i &=& \gamma_i \,[S(R)-3C(G)]^{-1}~,
\label{kappa}
\ee
which is consistent with the explicit result  \cite{kubo3}
in the minimal supersymmetric $SU(5)$ GUT.
Eq. (\ref{kappa}) together with eqs. (\ref{sumg}) and (\ref{cond2}) 
implies the sum rule (\ref{sumr})
so that the use of the sum rule 
in the intermediate steps of its derivation is
self-consistent. 
If one does not use the sum rule (\ref{sumr}), one finds
\be
\frac{d}{dt} \,(m^{2}_{i}+m^{2}_{j}+m^{2}_{k}
-M^{\dag} M) &=&
\sum_{r=i,j,k}\,\rho_{r p q}\rho^{r p q}(m^{2}_{r}+m^{2}_{p}+m^{2}_{q}
-M^{\dag} M)\,\frac{g^2}{16\pi^2}~,\\
& & \mbox{for}~~\{ i,j,k \}~~
\mbox{with}~~\rho_{ijk} \neq 0~.\nn
\ee
Therefore, if the sum rule is satisfied at some scale,
e.g. at the superstring scale, it remains for other scales.
If $S(R)-3C(G)=0$
(which means that $\beta_{g}^{(1)}=0$), eq. (\ref{kappa}) has no meaning.
In this finite case, the other $\beta$ functions also have to 
vanish as the consequence of the reduction equations (\ref{Yg2}), 
(\ref{redh}) and (\ref{redm}).
It is easy to see that the reduction solutions (\ref{Yg}) and (\ref{hY})
keep their form \cite{piguet1} and $[\beta_{m^2}]^{j}_{i}=0$
if the sum rule is satisfied. But the constant $\kappa_i$
remains undetermined.
In refs. \cite{jack1,kazakov1}, the symmetric choice
$m_{i}^{2}=m_{j}^{2}=m_{k}^{2}=(1/3)M M^{\dag}$
(see also ref. \cite{jones1}) has been made
to preserve higher-order finiteness of the SSB terms.

During the course of the derivation of the sum rule (\ref{sumr}),
we have made an interesting observation: 
The reduction solution for $h^{ijk}$ (\ref{hY}) means
that the so-called $A$ term is proportional to the
gaugino mass $M$, 
implying that  the
unwelcome superpartner 
contribution  to the electric dipole moment (EDM) 
is suppressed \cite{EDM,EDM2}.
Similarly, the reduced $B$-term is proportional 
to $\mu M$ (see the footnote 2).
If $\mu$ is real, the $B$-term and
the gaugino mass $M$ have the same
 CP phase, which  leads to  another suppression of EDM.

Let us next analyze how the relations
(\ref{sumr}) and (\ref{hY}) within the framework
of  supergravity can come about, where 
we do not necessarily think of supergravity as
an effective theory of superstrings for a while.
We begin by 
considering a  non-canonical K\"ahler potential
of the general form
\begin{eqnarray}
K = \tilde{K}(\Phi_a, \Phi^{*a}) + \sum_i K_i^i(\Phi_a, \Phi^{*a})
|\Phi^i|^2~, 
\label{K-non}
\end{eqnarray}
where $\Phi_a$'s and $\Phi_i$'s are 
chiral  superfields in the  hidden 
and visible sectors, 
respectively \footnote{In the following discussions, we adopt
the lazy notation that both
the chiral superfields and their scalar components are denoted
by the same symbol.}.
The basic assumptions to be made are:
\begin{enumerate}
\item Supersymmetry is broken by the $F$-term condensations 
($\langle F_a \rangle \neq 0$) of the   hidden sector fields
 $\Phi_a$.
\item The gaugino mass $M$ stems from
the gauge kinetic
function $f$ which depends  only on $\Phi_a$, i.e. $f=f(\Phi_a)$.
\item We consider only
those Yukawa couplings that have no field dependence.
\item The vacuum energy $V_0$ vanishes, i.e.,
\begin{eqnarray}
V_0 = \langle F_a F^b \tilde{K}^a_b \rangle - 3m_{3/2}^2 = 0~.
\label{V0}
\end{eqnarray}
\end{enumerate}
Under these assumptions, the SSB parameters can 
be written as \cite{SSB}
\begin{eqnarray}
 h^{ijk} &=& \langle F_a \rangle \langle (\,\tilde{K}
 - \ln(K_i^i K_j^j K_k^k)\,)^a \rangle\, Y^{ijk}~,
\label{A-non}\\
M &=& \langle F_a \rangle \langle (\,\ln \mbox{Re} f\,)^a \rangle~,
\label{M-non}\\
m^2_i &=& m_{3/2}^2 - \langle F_a \rangle \langle F^b \rangle
\langle (\ln(\,K_i^i\,))^a_b \rangle ~,
\label{m-non}
\end{eqnarray}
in the obvious notation. From eq. (\ref{hY}), 
upon  using eqs. (\ref{A-non}) and
(\ref{M-non}),  we  obtain the relation
\be
\langle F_a ~\rangle \langle (\ln \mbox{Re} f)^a  +
(\tilde{K} -  \ln(K_i^i K_j^j K_k^k)\,)^a ~\rangle =0~,
\ee
from which we deduce that, if
 $\langle F_a \rangle$'s are linear-independent, the constraint
\begin{eqnarray}
K_{(T)}(\Phi_a,\Phi^{*a}) &\equiv& 
\ln (K_i^i K_j^j K_k^k) \nn\\
&=&\tilde{K} + \ln \mbox{Re} f 
+\mbox{const.}~~\mbox{for all} ~~\{ i,j,k \}
~~\mbox{with} ~~ \rho_{ijk} \neq 0,
\label{tildeK}
\end{eqnarray} 
has to be satisfied.
Therefore,  there has to exist a definite relation between 
the K\"ahler potential $\tilde{K}$  in the hidden sector, 
the gauge kinetic function $f$ and the K\"ahler metric.
It is then straightforward to show that
the constraint (\ref{tildeK})  leads to the soft scalar-mass
sum rule (\ref{sumr}): 
\begin{eqnarray}
{\sum_l}^{\rho} m^2_l &=&3 m_{3/2}^{2}-
\langle F_a \rangle \langle F^b \rangle  \,
\langle K_{(T)\,b}^a \rangle =
 - \langle F_a \rangle \langle F^b \rangle 
\langle (\ln \mbox{Re} f)^a_b \rangle 
=M M^{\dagger}~,
\label{lhs}
\end{eqnarray}
where the use has been made of 
eqs. (\ref{m-non}), (\ref{M-non}), (\ref{tildeK}) and
the fact that
 Re$f$ is a direct sum of holomorphic functions of $\Phi_a$ and 
 $\Phi^{*\,a}$. 
We thus have arrived at the following 
generalized K\"ahler potential which
leads to (\ref{sumr}) and (\ref{hY}):
\begin{eqnarray}
 G &=& K + \ln|W|^2 ~,~ K_{(S)} (\Phi_a, \Phi^{*a}) 
=  -\ln(f(\Phi_a) +
\bar{f}(\Phi^{*a}))  ~,\nn\\
 K &=&  K_{(S)}(\Phi_a, \Phi^{*a}) 
+  K_{(T)}(\Phi_a, \Phi^{*a}) 
 + \sum_i K_i^i(\Phi_a, \Phi^{*a}) |\Phi^i|^2 ~.
\label{K-sol}
\end{eqnarray}
Given this K\" ahler potential, we next discuss symmetries behind.
One finds that there exist two types of symmetries:
The first one corresponds to  
the K\"ahler transformation together with  the chiral rotation of
the matter multiplets, 
\begin{eqnarray}
 \Phi_i &\to &  e^{{\cal M}_i} \Phi_i ~,~
 \Phi^{*i} \to e^{\bar{\cal M}_i} \Phi^{*i}~,\nn\\
K_i^i &\to & K_i^i  e^{-({\cal M}_i+\bar{\cal M}_i)}~,~
K_{(T)} \to K_{(T)} -{\cal M}-\bar{\cal M}~,\nn\\
f(\Phi_a) & \to & f(\Phi_a)~,~
W \to e^{{\cal M}}\, W~,
\label{chiral}
\end{eqnarray}
where ${\cal M}_i$ is a function of $\Phi_a$ and has to satisfy
the constraint
$\sum_{l} {}^{\rho} {\cal M}_l={\cal M}$ for all possible $\rho$'s
(the sum $\sum_{l} {}^{\rho}$ is defined in eq. (\ref{sumg})).
The second one is the invariance of
the K\"ahler metric $K_{(S)\,b}^{a}$ under the transformation
\begin{eqnarray}
f(\Phi_a) \to (af(\Phi_a)-ib)/(icf(\Phi_a)+d)~, 
\label{f-tr}
\end{eqnarray}
where $a$, $b$, $c$ and $d$ are integers satisfying $ad-bc=1$.
These symmetry properties are exactly what we are searching,
because we would like to understand the mass relations
(\ref{sumr}) and (\ref{hY}) in terms of  symmetries. From the 
transformation rules (\ref{chiral}) and (\ref{f-tr})
we see that it is likely for the symmetries
to be realized that the hidden sector fields are divided into
two types such that they enter either $K_{(S)}$ or $K_{(T)}$.
For 4D string models,
the symmetries (\ref{chiral}) and (\ref{f-tr}) indeed
appears as  the so-called target-space duality 
invariance (see refs. \cite{ferrara1,IL} and references therein)
and $S$-duality \cite{S-dual}, respectively.
Therefore, the sum rule (\ref{sumr}) and (\ref{hY})
might be naturally understood in superstring theories.

In the following discussion,  we restrict ourselves
to a certain class of the orbifold compactification 
and assume the existence
of a non-perturbative superpotential which breaks supersymmetry
and that the dilation $S$ and the overall modulus $T$ play
the dominant role for supersymmetry breaking.
So, $S$ and $T$ belong to the hidden sector.
It is known that the K{\" a}hler potential and 
the gauge kinetic function in this case 
assumes the  form
\begin{eqnarray}
&~& K=-\ln (S+S^*) -3 \ln (T+T^*)+\sum_i(T+T^*)^{n_i}|\Phi_i|^2~,~f=k S~,
\label{K-string}
\end{eqnarray}
where $n_i$ are (usually negative) integers and stand for modular 
weights, and  $k$ is
the Kac-Moody level  \cite{witten2}-\cite{DKL}.
The model possesses the $SL(2,Z)$ 
target-space duality invariance, and
under the duality transformation, the overall modulus $T$ transforms 
like \cite{ferrara3} 
\begin{eqnarray}
T \rightarrow (aT-ib)/(icT+d)~,
\end{eqnarray}
where $a$, $b$, $c$ and $d$ are integers satisfying $ad-bc=1$.
Chiral matter superfield $\Phi_i$ with
the modular weight $n_i$  transforms like  
\begin{eqnarray}
\Phi_i \rightarrow (icT+d)^{n_i}\Phi_i~
\label{transphi}
\end{eqnarray} 
so that the last term of $K$ remains invariant.
Comparing this with the transformation rule (\ref{chiral}), 
we see that ${\cal M}_i=n_i \,\ln (i c T+d)$, 
implying that the constraint $\sum_{l} {}^{\rho} {\cal M}_i = {\cal M}$
is satisfied only if $\sum_{l} {}^{\rho}\,n_l=$ const.
Since $K_i^i=(T+T^*)^{n_i}$, i.e., 
$K_{(T)}={\sum_{l} {}^{\rho}\,n_l\,}\ln (T+T^*)$,
the const. defined above has to be equal to $-3$ for  the K\" ahler
potential (\ref{K-string}) to belong to the class of (\ref{K-sol}).
The superpotential $W$, therefore,  transforms like
\begin{eqnarray}
W \rightarrow (icT+d)^{-3} W~,
\label{transW}
\end{eqnarray}
implying that  $W$ should have modular weight $-3$.
Since the modular weights are (usually) negative integers,
the matter chiral superfields $\Phi_i$'s appearing in 
non-vanishing cubic terms in $W$ ($\rho^{ijk} \neq 0$), have to 
have modular weight $-1$:
\be
n_i = n_j =n_k =-1~~\mbox{for}~~\{ i,j,k \}~~\mbox{with}~~
\rho^{ijk} \neq 0~, \ee
where we have assumed 
that the reduced Yukawa couplings $Y^{ijk}=
\rho^{ijk} g+\dots$ have vanishing modular weight.
Actually, within the framework of 
the orbifold models corresponding to the K\" ahler potential
(\ref{K-string}), we have \cite{BIM}
\be
M &=& \sqrt 3 m_{3/2} \sin \theta~,~
m_i^2 = m_{3/2}^2(1+n_i\cos^2 \theta) \label{stringm}~,\\
h^{ijk} &=& -\sqrt 3 Y^{ijk}m_{3/2}[\sin \theta 
+\cos \theta (3+n_i+n_j+n_k))]~, \label{stringm1}
\ee
where $\theta$ is the goldstino angle defined as 
$\tan \theta =\sqrt {K^S_S} F^S/(\sqrt {K^T_T} F^T)$.
Using the expressions above, one can explicitly check 
that the sum rule (\ref{sumr})
and (\ref{hY}) are satisfied \cite{BIM,BIM2,kim1}.

So far we have discussed only the soft scalar-mass
sum rule (\ref{sumr}) and (\ref{hY}).
Let us next briefly discuss on the individual reduction solution
(\ref{cond2}) itself. Comparing eqs. (\ref{cond2}) and (\ref{kappa}) 
with 
(\ref{stringm}), there has to exist
 a relation between $n_i$ and $\gamma_i$.
However, the allowed values of $n_i$ for the matter 
superfields are generally restricted \cite{IL,stringU}
so that the relation can not always be satisfied
for a given GYU model. To overcome this problem, 
we may consider
 multi-moduli cases, where  we have various modular weights and 
several goldstino angles as free parameters \cite{multiT,BIM2}. In
these cases, too, 
one can obtain the relation (\ref{hY}) and 
the sum rule (\ref{sumr})  \cite{kim1}. Another possibility is the D-term 
contribution
 to soft scalar-masses \cite{Dterm2} . 
This contribution can be written as \cite{Dterm} 
\begin{eqnarray}
{\cal D}_i &=& \sum_{A} q_i^A {\cal D}^A ~,~
{\cal D}^A = 2 g_A g_B (M_V^{-2})^{AB} \langle F^I \rangle
\langle F_J \rangle \langle (D^B)_I^J \rangle ~,
\label{DA}
\end{eqnarray}
where $q_i^A$ is the quantum number of $\Phi_i$ under the 
diagonal, broken gauge symmetries, 
$g_A$'s are gauge coupling constants, $(M_V^{-2})^{AB}$ is an inverse mass
matrix of massive gauge bosons and $D^A$ is the
corresponding $D$-term.
The indices $I,J$ run over all the chiral superfields.
Gauge invariance of the superpotential $W$ gives the constraint
$ {\sum_l}^\rho q_l^A = 0$,
which implies
$ {\sum_l}^\rho {\cal D}_l = 0$.
Therefore, the D-term
contribution does not affect the sum rule (\ref{sumr}),
but contributes to the individual soft scalar-masses.

Finally, we would like to emphasize
 that it is essential 
 for the sum rule (\ref{sumr}) and the relation (\ref{hY}) to be satisfied
that the Yukawa couplings  are reduced in favor of the gauge coupling $g$.
This is in sharp contrast to the case in
the effective supergravity; 
one can derive them 
 solely from the K\" ahler potential without 
explicitly knowing the superpotential. So, we can ask
ourselves why in GYU models  and in the certain superstring 
models  the same type of the soft-mass relations
are satisfied.
Ib\'a\~nez in
ref. \cite{ibanez1} gives an interpretation of the coincidence for the case
of finite GYU models in terms of the dilaton dominance 
and its relation to $N=4$ 
supersymmetry \footnote{As he points out,  the K\" ahler
potential (\ref{K-string}) indeed retains the $N=4$ structure,  
quite apart from the
fact that the matter fields generally are not  in the adjoint representation of 
the
gauge group. He argues that the coincidence of the  soft scalar-sum rule 
in finite GYU models should be  traced back  to its finiteness.}. 
 Here we argue
that not only in finite GYU models, but also in non-finite ones the
question could be  answered in terms of a sort of anomaly matching.

Given that
 the hidden sector fields $\Phi_a$'s  are supposed to
decouple at low energies (which is  an absolutely non-trivial assumption),
and that the duality transformation (\ref{chiral}) 
or (\ref{transphi}) acts as
$\Phi_a$-dependent scale transformations on the chiral matter
superfields, we may identify
the low-energy remnant of the duality invariance
with the  scale invariance of the effective
renormalizable field theory.
Needless to say that the transformation 
(\ref{transphi}) with $c=0$ defines
a global, common scaling for the  matter superfields in the untwisted sector,
and that 
the scaling of $W$ (see eq. (\ref{transW})) 
can be canceled 
 by an appropriate scaling of the superspace coordinates
(which, so to say, replaces the transformation of the hidden sector fields).
At this stage, we would like to recall that the duality invariance has an
anomaly  (absent at the string level), 
which is canceled by the Green-Schwarz mechanism and
the one-loop threshold corrections to gauge coupling 
$g$ \cite{ferrara1,IL}
in the effective supergravity.
The hidden sector fields $\Phi_a$'s play the basic 
role for this cancellation \cite{ferrara1,IL},
and in their absence at low energies, a definite
amount of the uncancelled anomaly remains.
In the class of the orbifold models, which we have considered above,
it is proportional  to
the one-loop coefficient $\beta_{g}^{(1)}$ 
of the $\beta$ function of  $g$ \cite{ferrara1,IL}.
So, if the scale invariance should be identified with the low-energy
remnant of the duality invariance, 
its anomaly, too, should be controlled by the 
$\beta$ function of the gauge coupling $g$.
However, the $\beta$ functions
of the Yukawa couplings also
contribute to the scale invariance anomaly
at the renormalizable level; unless
they
are reduced  in favor of the gauge coupling $g$
so that there exists only one $\beta$ function in a  theory
which dictates the anomalous scaling behavior.
Note that the reduction equation (\ref{Yg2}) defines a set of ordinary
differential equations of first order, and so  
the general solutions contain
a set of integration constants.
But
the requirement that 
the solution is consistent with perturbative renormalizability
is usually sufficient
to obtain a unique solution \cite{zim1}. 
Moreover, only this power series solution 
(\ref{Yg}) is regular in the sense
that it has the well-defined  
$ \beta_{g}^{(1)} \to 0 $ limit, the scale
invariance limit \cite{LPS} 
(where we assume we could vary $\beta_{g}^{(1)}$
 smoothly).
The other solutions  have an essential singularity
in this limit. Since we do not expect such a singularity
at the effective supergravity level,
the power series solution is singled out:
Shortly, GYU is a consequence of the anomaly matching.

We believe that the simplicity and the universality
of the soft mass relations in GYU models and 
the coincidence with their
corresponding relations in superstring models have 
a deep meaning, and hope that our interpretation of it
will put us toward a more complete understanding
of the relation between gauge-Yukawa unified
and superstring theories.

\vspace{0.7cm}
\noindent
We thank D. Suematsu and G. Zoupanos for
useful discussions.

\end{document}